\documentclass[pra,twocolumn,showpacs,superscriptaddress]{revtex4-1}

\usepackage{multirow,eurosym,amssymb,amsfonts,amsmath,graphicx,color,bm,float,verbatim}

\newcommand{\bra}[1]{\langle #1|}
\newcommand{\ket}[1]{|#1\rangle}

\def\be{\begin{equation}}
\def\ee{\end{equation}}

\def\bsplit{\begin{split}}
\def\nsplit{\end{split}}

\begin{document}

\title{Quantum metrology for non-linear phase shifts with entangled coherent states}

\date{\today}

\author{Jaewoo Joo}
\affiliation{Quantum Information Science, School of Physics and Astronomy, University of Leeds, Leeds LS2 9JT, U.K.}

\author{Kimin Park}
\affiliation{Center for Macroscopic Quantum Control and Department of Physics and Astronomy, Seoul National University,
Seoul 151-747, Korea}

\author{Hyunseok Jeong}
\affiliation{Center for Macroscopic Quantum Control and Department of Physics and Astronomy, Seoul National University,
Seoul 151-747, Korea}

\author{William J. Munro}
\affiliation{NTT Basic Research Laboratories, NTT Corporation, 3-1 Morinosato-Wakamiya, Atsugi-shi, Kanagawa 243-0198, Japan}
\affiliation{ National Institute of Informatics 2-1-2 Hitotsubashi, Chiyoda-ku, Tokyo 101-8430, Japan}
\affiliation{Quantum Information Science, School of Physics and Astronomy, University of Leeds, Leeds LS2 9JT, U.K.}

\author{Kae Nemoto}
\affiliation{ National Institute of Informatics 2-1-2 Hitotsubashi, Chiyoda-ku, Tokyo 101-8430, Japan}

\author{Timothy P. Spiller}
\affiliation{Quantum Information Science, School of Physics and
Astronomy, University of Leeds, Leeds LS2 9JT, U.K.}

\begin{abstract}
We investigate the phase enhancement of quantum states subject to non-linear phase shifts. The optimal phase estimation of even entangled coherent states (ECSs) are shown to be better than that of NOON states with the same average particle number $\langle n \rangle$ and non-linearity exponent $k$. We investigate the creation of an approximate entangled coherent states (AECS) from a photon subtracted squeezed vacuum with current optical technology methods and show that a pure AECS is even better than an even ECS for large $\langle n \rangle$. Finally we examine the simple but physically relevant cases of loss in the non-linear interferometer for a fixed average photon number $\langle n \rangle$.
\end{abstract}


\maketitle
\section{Introduction}
\label{sec1} 

Quantum metrology is a research field that examines the characteristic fundamental properties of measurements under the laws of quantum mechanics \cite{NOON05,NOON01}. The ultimate goal of this is to achieve measurements  at the information theoretical bounds allowed by the laws of quantum mechanics, far beyond their classical counterparts. For optical system the classic and extremely well studied example are NOON states \cite{NOON05,NOON01,NOON02,NOON03,NOON04,NOON06} whose performance allows them to measure linear phase shifts at the Heisenberg limit. Several
theoretical studies have recently investigated the role of non-linearity to help improve the limits of phase enhancement in linear systems  \cite{gerry07,Nonlinear01,Nonlinear02,Nonlinear03,Mitchell10} and the first demonstration of this so-called super-Heisenberg scaling has been shown  \cite{Demo11}. The particle-loss and decoherence mechanisms are, however, not fully explored in combined linear and non-linear interferometers, even theoretically \cite{WalmsleyNature10,WalmsleyPRL09,WalmsleyPRA09,Escher11}.

It is not only NOON states that allow linear phase measurements at the Heisenberg limit.  Entangled coherent states (ECSs)  \cite{Barryarxiv,gerry95,Barryarxiv1,Barryarxiv3,Enk01,Jeong01,munro02,Peter05} are also able to do this \cite{Gerry02,Gerry01,Gerry03} and can outperform that of NOON states in the region of very modest particle numbers with a linear phase operation \cite{Joo11,NewArxiv12}. An important case of entangled coherent states are the two-mode path entangled states, a state analogous to a NOON state, but with one of two modes containing a coherent state rather than a Fock state \cite{Gerry01}. This particular ECS can be represented as a superposition of NOON states with different
photon numbers \cite{Gerry02}. Using linear optical elements, the phase sensitivity of ECSs outperforms that of NOON and bat states \cite{Jess10}, both without\cite{Gerry02,Gerry01,Gerry03} and with losses \cite{Joo11}, because coherent states maintain their properties in the presence of loss. Given all this recent work, a natural question arises regarding comparison of the phase enhancement for ECSs, compared to NOON and other states, in the case of non-linear phase shifts \cite{gerry07}.

In this paper we are going to  investigate the non-linear phase enhancement resulting from a generalized non-linearity characterized by an exponent $k$ ($k>0$) on  four quantum states: NOON, even and odd ECSs and an approximated ECS (AECS). The AECS is created from a photon subtracted squeezed state and experimentally feasible to realize \cite{photon_sub01,photon_sub02}. Potential enhancements will be quantified with the quantum Fisher information \cite{Braunstein94,Braunstein96}. To begin we will consider a two-mode pure state $|\psi \rangle_{12}$ and a generalized non-linear phase shifter $U(\phi,k)$ given by
\begin{eqnarray}
U(\phi,k) &=& {\rm e}^{i \phi (a^{\dag}_2 a_2)^k},
\label{eq:Phase01}
\end{eqnarray}
where $a^{\dag}_i$ ($a_i$) is a creation (annihilation) operator in spatial mode $i$ \cite{BillPRA2010} (see the details in Section \ref{sec2-1}). The exponent $k$ represents the order of the non-linearity. For example, $k = 1$ corresponds to a linear phase shift on the state, $k =2$ a Kerr phase shift and $k \neq 2$ gives a more general non-linear effect in a phase operation. When the generalized phase operation $U(\phi,k)$ is applied to mode 2 of $|\psi \rangle_{12}$, the resultant state is equal to
\begin{eqnarray}
&& |\psi^k (\phi) \rangle_{12} = (\openone \otimes U(\phi,k) )
|\psi \rangle_{12}.
\end{eqnarray}
Now according to phase estimation theory \cite{Braunstein94,Braunstein96}, the phase uncertainty is bounded by the quantum Fisher information
\begin{eqnarray}
\delta \phi & \ge & {1\over \sqrt{F^Q}}, 
\label{eq:Fisher01}
\end{eqnarray}
where the value of $F^Q$ for pure states is simply given by
\begin{eqnarray}
&&F_{}^{Q} = 4\Big[ \langle \tilde{\psi}^k  |\tilde{\psi}^k \rangle - |\langle \tilde{\psi}^k |\psi^k (\phi) \rangle |^2 \Big] = 4 (\Delta n^k)^2, 
\label{eq:Fisher02}
\end{eqnarray}
with $| \tilde{\psi}^k \rangle = \partial |\psi^k (\phi) \rangle/\partial \phi$ and $(\Delta n^k )^2 = \langle (n^k)^2 \rangle - \langle n^k \rangle^2$ ($\langle n^k \rangle= {}_{12} \langle \psi | ( a^{\dag}_2 a_2 )^k |\psi\rangle_{12}$). It is important to note that $\langle n^1 \rangle$ denotes an average (or mean) photon number.

To allow a fair comparison of the phase sensitivity among the various different quantum states under consideration, we will use the same average photon number in one of two modes as the physical resource count for the states \cite{Gerry02,BillPRA2010,gerry2000,gerry2002}. For pure states, we shall demonstrate an inequality for the sensitivity among the three states: NOON (least sensitive), odd ECS and even ECS (most sensitive). Furthermore we will show that  in the limit of large $\langle n^1 \rangle$ the AECS slightly better than the other three states.  We shall also consider a small amount of loss for the dispersive and non-linear interferometer arm, because the non-linear medium providing the phase operation will generally also provide a scattering effect (e.g., particle losses) \cite{ref1:keq2}. We shall demonstrate that, analogous to the linear case ($k=1$) \cite{Joo11,NewArxiv12}, the phase enhancement of ECSs still outperforms that of NOON states, even for non-linear cases ($k\neq1$).

For a physical realization, it is known that the ideal ECS can be generated by mixing through a 50:50 beam-splitter (BS) a coherent state and a coherent state
superposition (CSS) \cite{Peter05}, given by
\begin{eqnarray}
|CSS_{\pm} (\alpha) \rangle = {N}^{\pm}_{\alpha} (|\alpha \rangle \pm |{\rm -} \alpha \rangle)
\end{eqnarray}
where $|\alpha \rangle$ is a coherent state with amplitude $\alpha$ and ${N}^{\pm}_{\alpha} = 1 / \sqrt{2(1 \pm {\rm e}^{-2|\alpha|^2})}$. Since the CSS with small $\alpha$ has been already demonstrated in experiments \cite{Grangier09,Polzik06,Furusawa08,Knill10}, the scheme of building AECSs with a modest photon number may be experimentally feasible with very high fidelity in the near future \cite{LundPRA2004}. Several experiments have demonstrated that non-linear phase operations can be realized in various set-ups. For example, self-Kerr phase modulation ($k=2$) has been measured as a function of electric field amplitudes in waters, fibers, Nitrobenzene, Rydberg states, etc. \cite{ref1:keq2,ref2:keq2,ref3:keq2}. Notably, the phase shift dependent upon the applied field clearly follows theoretical predictions in the case of a Rydberg electro-magnetically induced transparency medium (see in Fig.~3 in Ref.~\cite{ref1:keq2}).

The remainder of this paper is organized as follows. In Sec. \ref{sec2-1}, the mathematical formalism for generalized phase shifters is described. The phase enhancements of pure ECSs is discussed in the case of ideal preparation in Sec. \ref{sec2-2}. A feasible approach to implementing the AECS is given in Sec.~\ref{sec2-3}. In Sec.~\ref{sec3} we also investigate the effects of loss on the phase enhancement behavior. Finally, in Sec.~\ref{sec4} we summarize and discuss our results.

\section{Optimal phase estimation using nonlinearity in pure states} 
\label{sec2-1}
Let us first discuss the validity of Eq.~(\ref{eq:Phase01}) for general $k$. 
The generalised phase operation is formed by
\begin{eqnarray}
U(\phi) = \exp \Big[ i \hat{H} (\phi) \,t / \hbar \Big],
\end{eqnarray}
where the total Hamiltonian is equal to
\begin{eqnarray}
\hat{H} (\phi) &=&  \hat{H}_0 + \hat{H}_{int} (\phi),
\label{Hamiltonian0}
\end{eqnarray}
consisting of the unperturbed Hamiltonian represented by
\begin{eqnarray}
\hat{H}_0 &=& \int d^3 r \Big[ \frac{1}{2\mu_0} |\hat{B}|^2 + \frac{\epsilon_0 }{2} |\hat{E}|^2 \Big] = \hbar \omega \Big( a^{\dag} a + \frac{1}{2}\Big), ~~\label{Hamiltonian00}
\end{eqnarray}
for mode frequency $\omega$ and an interaction Hamiltonian given by expanding the polarization of the non-linear medium
\begin{eqnarray}
\hat{H}_{int} (\phi) &=& \int d^3 r \, \Big[ \hat{E} \cdot \hat{P} \Big] = \int d^3 r \, \hat{E} \Big[ \sum_{j=1}^{\infty} \frac{\chi^{(j)}}{j+1} (\hat{E})^j \Big], ~~~~ \label{Hamiltonian01}
\end{eqnarray}
where $\chi^{(j)}$ is the $j$-th order susceptibility of the medium \cite{Drummond}. A single-mode electric field is given by
\begin{eqnarray}
\hat{E} = i \sqrt{\frac{\hbar \omega}{2 \epsilon_0}} \Big( a u(r) - a^{\dag} u^*(r) \Big), \label{E01}
\end{eqnarray}
where $u(r)$ is the mode function. Due to the lack of phase matching, the single-mode assumption and the rotating wave approximation, we may
neglect the terms in $\chi^{(2x)}$ ($x$: positive integer) \cite{Drummond}, and then,
\begin{eqnarray}
U (\phi,k) = \exp \Big[ i \omega t  (a^{\dag} a + \frac{1}{2})\Big] \prod_{k=1}^{\infty} \exp \Big[ i \phi^{(k)} \Big( a^{\dag} a \Big)^k \Big],~~~~~
\end{eqnarray}
where the phase parameter of the non-linearity $k$ is
\begin{eqnarray}
 \phi^{(k)} &=& t \int d^3 r  \sum_{x=1}^{\infty} {\cal{F}} \Big( \chi^{(2x-1)} \Big) .
\label{phik}
\end{eqnarray}
We note that $ {\cal{F}} ( \chi^{(2x-1)} )$ is a function of $\chi^{(2x-1)}$. Therefore, the expression of the non-linear phase operation in Eq.~(\ref{eq:Phase01}) is appropriate for fixed $k$.

A well-known example of a non-linear phase operation is given by the Kerr interaction for $k=2$  \cite{Gerry02,Gerry01,Haus89}. In an interaction picture that removes the linear dynamical phase, the non-linear component is
\begin{eqnarray}
U(\phi,2) = \exp \Big[ i \phi^{(2)} \Big( a^{\dag} a \Big)^2 \Big],
\end{eqnarray}
with
\begin{eqnarray}
 \phi^{(2)} &=& t \int d^3 r \Big( \frac{3}{2} \chi^{(3)}+ 5 \chi^{(5)} \Big),
\end{eqnarray}
where the interaction Hamiltonian is truncated after the fifth-order susceptibility $\chi^{(5)}$.

\subsection{Ideal (theoretical) cases}
\label{sec2-2}

\begin{figure}[t]
\center
\includegraphics[width=270px]{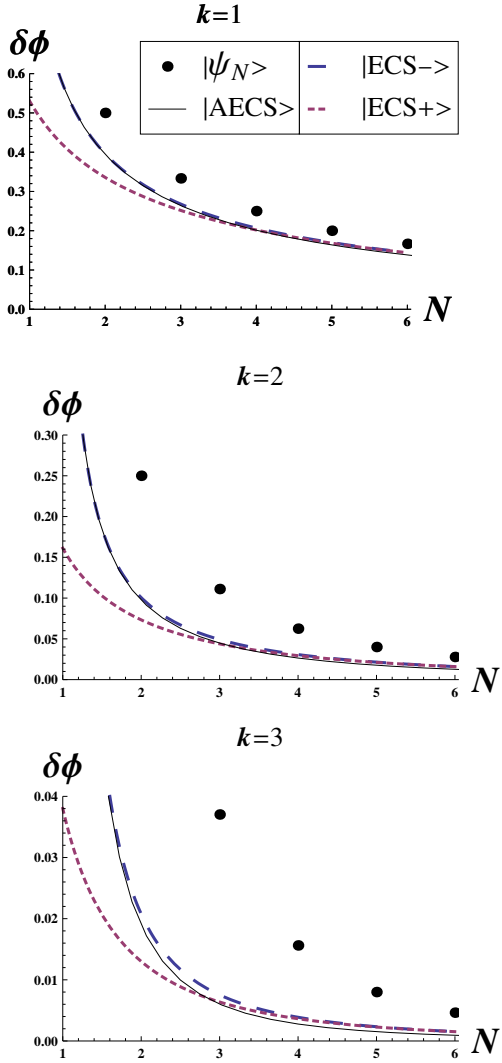}
\caption{ (color online). The plots show the inequality $\delta
\phi_{N^k} \ge \delta \phi_{E^k_{-}} \ge \delta \phi_{E^k_{+}}$
with respect to $N=2\langle n^1_N \rangle =\langle n^1_{E_{\pm}}
\rangle$ ($k=1,2,3$). \label{fig:Ideal01} }
\end{figure}

We now need to review and calculate the phase enhancements using the quantum
Fisher information for pure (no loss) cases of the NOON state and ECSs. The NOON state is defined by \cite{NOON03}
\begin{eqnarray}
|\psi_{N}\rangle_{12} = \frac{1}{\sqrt{2}} (|N\rangle_1 |0\rangle_2 + |0\rangle_1 |N\rangle_2),
\end{eqnarray}
where $|N\rangle$ is a number state with photon number $N$. After a generalized phase shifter $U(\phi,k)$ is applied in mode 2, the resulting state is  given by $|\psi_{N}^{k} \rangle_{12} =\Big( \openone \otimes U(\phi,k) \Big)|\psi_{N} \rangle_{12}$.  From Eq.~(\ref{eq:Fisher02}), the quantum Fisher information of the pure NOON states with a non-linearity of order $k$ is given by
$F_{N^{k}}^{Q}=1/N^{2k}$ and 
\begin{eqnarray}
\delta \phi_{N^k} > {1 \over N^{k}}. 
\label{eq:FINOON01}
\end{eqnarray}
Similarly for the even and odd ECSs defined by
\begin{eqnarray}
&&|ECS_{\pm} (\alpha_{\pm}) \rangle = {\cal{N}}^{\pm}_{\alpha_{\pm}} \Big[|\alpha_{\pm} \rangle_{1}
|0\rangle_2 \pm |0 \rangle_1 | \alpha_{\pm} \rangle_2 \Big],
\label{eq:Coherent01}
\end{eqnarray}
with amplitude $\alpha_{+}$ ($\alpha_{-}$) for even (odd) ECS and ${\cal{N}}^{\pm}_{\alpha_{\pm}} = 1 / \sqrt{2(1 \pm {\rm e}^{-|\alpha_{\pm}|^2})}$. After the phase shifter $U(\phi,k)$ is performed in mode 2, we find that the resulting state is given by
\begin{eqnarray}
|ECS^{k}_{\pm} (\alpha_{\pm}, \phi) \rangle_{12} = \Big(\openone \otimes U(\phi,k) \Big) |ECS_{\pm} (\alpha_{\pm}) \rangle_{12}.
\end{eqnarray}
The quantum Fisher information is then given by
\begin{eqnarray}
&&F_{E^{k}_{\pm}}^{Q} =  4 f_{\alpha_{\pm}} \left[ \sum_{n=0}^{\infty}  { n^{2k} (\alpha_{\pm})^{2n} \over n!}  - f_{\alpha_{\pm}} \left(
\sum_{n=0}^{\infty}  { n^k (\alpha_{\pm})^{2n} \over n!} \right)^2 \right]. \nonumber \\
\label{eq:QFECS}
\end{eqnarray}
for $f_{\alpha_{\pm}} = {\rm e}^{- {|\alpha_{\pm}|^2 }} ({\cal{N}}^{\pm}_{\alpha_{\pm}})^2$ and
\begin{eqnarray}
\delta \phi_{E^{k}_{\pm}} > {1 \over \sqrt{F_{E^{k}_{\pm}}^{Q}}}.
\label{eq:FIECS01}
\end{eqnarray}

\begin{figure}[b]
\centering
\includegraphics[width=220px]{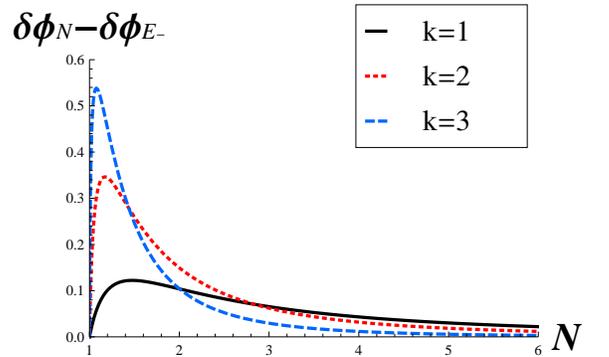}
\caption{ (color online). Difference of the optimal phase estimation between $|\psi_N \rangle$ and $|ECS^{k}_{-} (\alpha_-) \rangle$ such as $\delta \phi_{N^k} - \delta \phi_{E^k_-}$ for $k=1,2,3$.} \label{fig:Delta_keq123}
\end{figure}

In order to compare the phase sensitivity of the different states, we take into account the same {\it average} particle
number of the states \cite{Gerry02,BillPRA2010,gerry2000,gerry2002} in an arm
\begin{eqnarray}
\langle n^1_N \rangle = \langle n^1_{E_{\pm}} \rangle = {N \over 2} = \big({\cal{N}}^{\pm}_{\alpha_{\pm}}\big)^2 \cdot |\alpha_{\pm}|^2, \label{APN01}
\end{eqnarray}
where $\langle n^1_{E_{\pm}} \rangle= \langle ECS_{\pm} (\alpha_{\pm}) | a^{\dag}_2 a_2 |ECS_{\pm} (\alpha_{\pm}) \rangle$ and in general $\alpha_{+} \neq \alpha_{-}$. In Fig.~\ref{fig:Ideal01}, the values of optimal phase estimation are plotted for the three quantum states and satisfied with the inequality
\begin{eqnarray}
\delta \phi_{N^k} & \ge & \delta \phi_{E^k_{-}} \ge \delta \phi_{E^k_{+}}
\label{inequality01}
\end{eqnarray}
for any $N$ and $k$. The first inequality $\delta \phi_{N^k} \ge \delta \phi_{E^k_{-}}$ shows the pattern of the difference $\delta \phi_{N^k} - \delta \phi_{E^k_{-}}$ for $k=1,2,3$ in Fig.~\ref{fig:Delta_keq123}. Note that $\delta \phi_{E^k_{\pm}}$ approaches to $\delta \phi_{N^k}$ because of $|ECS_{\pm} (\alpha) \rangle \approx |\psi_{N}\rangle$ for larger $N$ and that $\delta \phi_{E^k_{\pm}}$ is a continuous value while $\delta \phi_{N^k}$ exists discretely due to integer $N$.

\subsection{Preparation of an approximate ECS}
\label{sec2-3}

\begin{figure}[t]
\hspace{-1.7cm}
\includegraphics[angle=-90,width=210px]{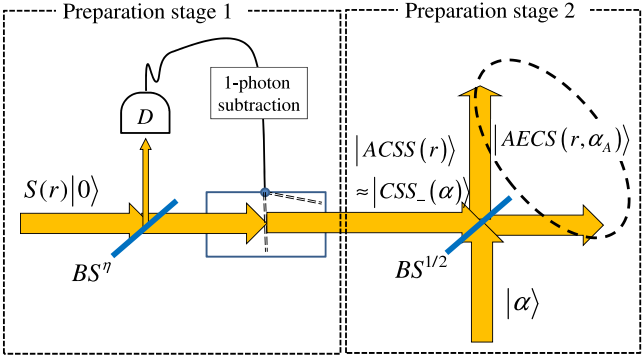}
\vspace{-0.7cm} \caption{ (Color online) Schematics of generating
$|AECS (r, \alpha_A) \rangle$ from a squeezed vacuum $S(r)|0\rangle$ and a coherent state $|\alpha \rangle$. After single-photon subtraction at stage 1, the state $\ket{ ACSS (r) }$ is very similar to $|CSS_- (\alpha) \rangle$. The coherent state in stage 2 has the same amplitude of the old CSS ($BS^{\eta}$ and
$D$ are an unbalanced BS with transmission rate $\eta$ and a single-photon detector).
\label{fig:Prepare01}}
\end{figure}

We now present an optical set-ups to create an AECS with a very high fidelity to an odd ECS, based on current optical technology methods. We also compare the phase enhancement of the AECS with the other states. As shown in Fig.~\ref{fig:Prepare01}, two steps are required for AECS preparation. First, we create a photon-subtracted quantum state, with high fidelity to an odd CSS, given by the scheme in Ref.~\cite{LundPRA2004}. The generation of squeezed vacuum states can
be given by degenerate parametric down-conversion utilizing non-linearity \cite{Peter05} and a series of experimental results shows that modest strengths of squeezing through second-harmonic generation is achievable with the current technology \cite{Schnabel10,Vahlbruch,Chelkowski}. The scheme of single-photon subtraction through an unbalanced BS from a squeezed vacuum has been already demonstrated \cite{photon_sub01,photon_sub02}. Finally, generation of the odd ECS follows from the well known technique of mixing a traveling CSS with a controlled coherent state through a 50:50 beam-splitter (BS).

In Fig.~\ref{fig:Prepare01}, the schematics shows how to generate $|AECS (r, \alpha_A) \rangle$ from a squeezed vacuum $S(r)|0\rangle$ and a coherent state $|\alpha \rangle$ ($S(r)=\exp\left[-{r\over 2} \left( a^2-(a^{\dag})^2\right)\right]$ and $r$ is a squeezing parameter). It was shown that the fidelity between a squeezed single photon state $S(r)\ket{1}$ and an odd CSS $|CSS_{-} (\alpha) \rangle$ with small $\alpha$ is extremely high \cite{LundPRA2004} and that a photon-subtracted squeezed vacuum state $a S(r)\ket{0}$ is identical to $S(r)\ket{1}$ \cite{Jeong05}. We begin by preparing a squeezed vacuum $S(r)|0\rangle$ and then performing single-photon subtraction by using $BS^{\eta}$ ($\eta$: transmission rate) and a single-photon detector. The resultant state is called an ACSS possessing a very high fidelity compared with the ideal odd CSS. In detail, when a single photon is detected, the resultant state $\ket{ ACSS (r_0)}= a S(r_0)\ket{0}$ ($r_0 = {\rm arcsinh} [2\alpha_0 /3]$) is given by
\begin{eqnarray}
\label{eq:normalized} \ket{ ACSS (r_0) } = f_r \sum_{k=0}^{\infty}\frac{\sqrt{(2k+1)!}}{2^k \cdot k!}(\tanh r_0)^k\ket{2k+1},~~~~~ 
\end{eqnarray}
for $f_r =(1-\tanh^2 r_0)^{3/4}$ with the maximized fidelity between $\ket{ ACSS (r_0) }$ and $|CSS_{-}(\alpha_0) \rangle$.

\begin{figure}[b]
\hspace{-0.5cm}
\includegraphics[width=250px]{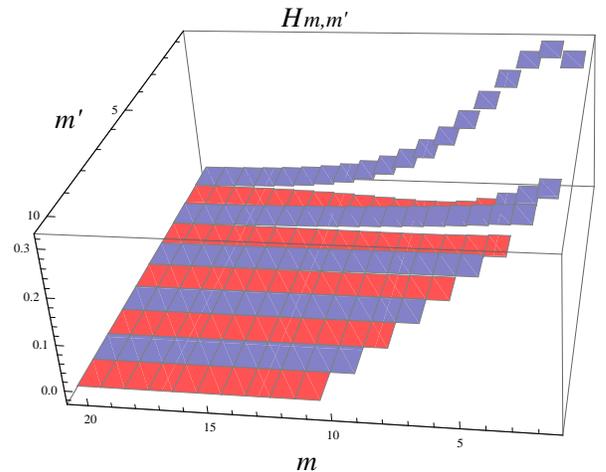}
\caption{ (color online). The coefficient $H(m,m')$ of $|AECS (r_0,\alpha_A)\rangle$ ($\alpha_A \approx 1.9807$). It clearly shows that the major contribution of the NOON state and the minor of $m$ and $m'$ states. The blue (red) colour indicates a positive (negative) value. 
\label{fig:Hmn} }
\end{figure}

In the second stage, the odd AECS can be built with the generated ACSS $\ket{ ACSS (r_0)}_1$ mixed with an extra coherent state $|\alpha_0 \rangle_2$ through an 50:50 BS. The state is written as
\begin{eqnarray}
\label{eq:AECS01}
\ket{{AECS} (r_0,\alpha_A)}&=&\sum_{m=1}^\infty \sum_{m'=0}^{m-1} \Big[ H_{m,m'} (\ket{m}\ket{m'} - \ket{m'}\ket{m}) \Big], \nonumber \\ 
&\approx& \ket{ECS_- (\alpha_-)},
\end{eqnarray}
where $H_{m,m'}$ is the coefficient of the state in Fock basis ($\alpha_A = \sqrt{2} \alpha_0$). As shown in Fig.~\ref{fig:Hmn}, the resultant state $\ket{{AECS} (r_0,\alpha_A)}$ is approximately equivalent to the desired odd ECS $|ECS_{-} (\alpha_-) \rangle$ with high fidelity ($\approx 0.975$) if $\alpha_- = \alpha_A \approx 1.9807$. The state $\ket{{AECS} (r_0,\alpha_A)}$ consists of the ideal ECS ($m'=0$) and the unbalanced photon states called $m$ and $m'$ states ($m' \neq 0$) \cite{mandmstate}. In other words, the outcome state contains a superposition of NOON states for $m'=0$ while it also includes the states possessing unbalanced photon numbers in both modes for $m' \neq 0$. Then, after the generalized phase shifter $U(\phi,k)$ in mode 2, we can estimate the phase enhancement of the final state given by $\ket{{AECS}^{k} (r_0, \alpha_A,\phi)}  = (\openone \otimes U(\phi,k) )\ket{{AECS} (r_0,\alpha_A)}$.

\begin{figure}[htb]
\centering
\includegraphics[width=250px]{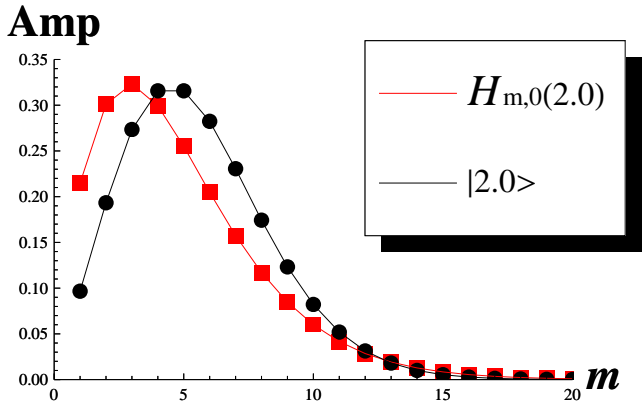}
\caption{ (color online). The amplitude of $H_{m,0} (\alpha_A)$ for $\alpha_A =2.0$ and that of coherent state $|2.0\rangle$ with respect to photon number $m$. Both states contain the same average photon number. It implies that the AECS reaches the peak amplitude in smaller $m$ and has a longer tail than the coherent state. 
\label{fig:NOONF} }
\end{figure}

For the phase enhancement of the AECS compared with ideal ECSs, the value of $\langle n^1_{E_{A}} \rangle = \bra{AECS } a^{\dag}_2 a_2 \ket{AECS }$ should be $N/2$ in
Eq.~(\ref{APN01}). As shown in Fig.~\ref{fig:Ideal01}, it is true that the results from AECS are very close to those from ideal odd ECSs for $k=1,2$ but slightly better than those of the other states at $k=3$, for the modest number of $N$ (see thin black lines in Fig.~\ref{fig:Ideal01}). This is because the detailed shape of the AECS is slightly different from the ideal odd ECS. In Fig.~\ref{fig:NOONF}, this advantage of phase enhancement can be explained by the fact that the distribution of $H(m,0)$ is narrower but has a longer tail in the AECS, compared to a coherent state $|2.0\rangle$, indicating the photon distribution of $|ECS_{\pm} (\alpha_\pm) \rangle$, for the same average photon number. The $m$ and $m'$ states might provide
a minor contribution of phase enhancement in the AECS.

\section{For small loss cases in a non-linear phase operation}
\label{sec3}

The phenomenon of imperfect phase operations may lead to small particle losses in the arm. For the comparison of phase enhancement in lossy states, we choose the fixed average
photon number of the states such as
\begin{eqnarray}
\label{eq:Fixed_Ave01} \langle n^1_{N} \rangle = \langle n^1_{E_{+}} \rangle = \langle n^1_{E_{-}} \rangle = \langle n^1_{E_{A}} \rangle = 2.0,
\end{eqnarray}
which implies $N = 4$. For example, $\ket{{AECS} (r_0,2.0}$ and $\ket{{ECS}_- (1.9807)}$ have the same average photon number such as $\langle n^1_{E_{-}} \rangle = \langle n^1_{E_{A}} \rangle = 2$.

Adding a BS with vacuum input can mimic this lossy condition in the dispersive interferometer arm after the non-linear phase shift ($T$: transmission rate of the BS) \cite{Escher11}. Here, we examine the phase enhancement of mixed states by generalising the results of Ref.~\cite{Escher11} such as
\begin{eqnarray}
\label{eq:Fixed_Ave02} && \delta \phi \ge \frac{1}{\sqrt{F^{Q}}} \ge  \frac{1}{\sqrt{C^{Q}_k}} \,,
\end{eqnarray}
where $C^{Q}_k = 4 \Big( \langle H^k_1 \rangle - \langle H^k_2 \rangle^2 \Big)$ for any $k$ \cite{CQ}. In particular, this equation shows an excellent match with the exact value of the quantum Fisher information in the small loss region ($T\approx 1$) \cite{Escher11}. For $k=1$, the bound $C^Q$ is given by
\begin{eqnarray}
\label{eq:CQkEq1} C^Q_1 = 4 \Big[ T^2 \Big( \langle n^2 \rangle - \langle n \rangle^2 \Big) +T(1-T) \langle n \rangle \Big],
\end{eqnarray}
and for $k=2$,
\begin{widetext}
\begin{eqnarray}
\label{eq:CQkEq2}
C^Q_2 &=& 4\Big[ T^4 \,\langle n^4 \rangle + 6 T^3 (1-T) \, \langle n^3\rangle +  T^2 (1-T)(3-11 T) \, \langle n^2 \rangle + T(1-T)(1- 6T + 6 T^2) \langle n \rangle \nonumber \\
&& - \Big( T^4 \langle n^2 \rangle^2 + 2 T^3 (1-T) \langle n \rangle \langle n^2 \rangle + T^2 (1-T)^2 \langle n \rangle^2 \Big) \Big].
\end{eqnarray}
\end{widetext}

\begin{figure}[htb]
\hspace{-0.7cm}
\includegraphics[width=270px]{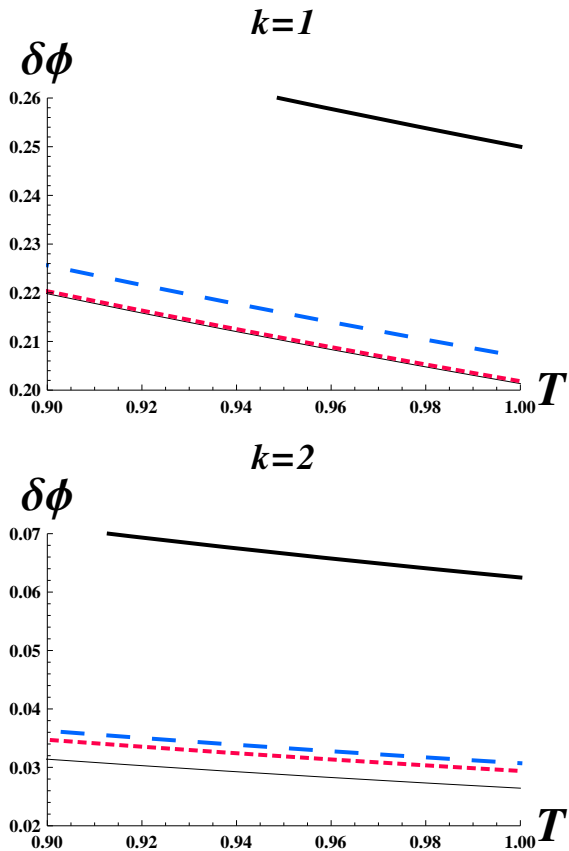}
\caption{ The phase sensitivity with small lossy conditions for $k=1,2$ ($T$: transmission rate of the BS mimicking photon losses). The thick solid line is for NOON states and the thin solid line is for AECS which is very similar to dashed lines for even (short-dashed) and odd (long-dashed) ECSs. 
\label{fig:Lossy} }
\end{figure}

As shown in Fig. \ref{fig:Lossy}, the ECSs including AECS significantly outperform NOON states (thick solid lines) for $k=1,2$ in a small photon-loss window. For NOON, even and odd ECS and AECSs, the expectation values in Eqs.~(\ref{eq:CQkEq1}) and (\ref{eq:CQkEq2}) are given by
\begin{eqnarray}
\langle n^k \rangle &=& {N^{k}\big/ 2},\\
\langle n^k_{E_{\pm}} \rangle &=& f_{\alpha_{\pm}}
\sum_{n=0}^{\infty}  { n^k (\alpha_{\pm})^{2n} \over n!},\\
\langle n^k_{E_{A}} \rangle &=& \bra{AECS (r,\alpha_A)} (a^{\dag}_2 a_2)^k \ket{AECS (r,\alpha_A)}. \nonumber \\
&=& \sum_{m=1}^\infty \sum_{m'=0}^{m-1} \Big[  \Big( H_{m,m'}\Big)^2 \Big( m^{k} + (m')^{k} \Big) \Big].
\end{eqnarray}
Therefore, these results show that ECSs still outperform the phase enhancement achieved by NOON states in the region of small losses after the non-linear phase operation ($k=2$).

\section{Summary and Remarks}
\label{sec4} 

In summary, we have analyzed phase enhancement of ECSs for non-linear phase shifts, using quantum Fisher information to quantify the results. As shown in linear optical elements \cite{Joo11,NewArxiv12}, the phase sensitivity of ECSs outperforms that of NOON states for modest average photon numbers, converging to the limit of NOON states for large average photon numbers. We have presented the form of generalized (non-linear) phase operations, in terms of the power of number operators, and obtained an inequality for the phase enhancement of NOON states and odd and even ECSs, all with respect to the same average photon number as a physical resource. We have also investigated the feasibility of creating AECS in optical set-ups based on current technology, and examined its phase sensitivity. Finally, we have shown that the behavior of the phase sensitivity for ECSs significantly outperforms that for NOON states, for the $k=2$ non-linear example in the presence of small losses.

The fidelity between the odd ECS $|ECS_{-} (\alpha_-) \rangle$ and AECS $\ket{{AECS} (r_0,\alpha_A)}$ is very high, but in general the AECS has more degrees of freedom with $r$ and $\alpha$ (see Fig. \ref{fig:Prepare01}). Thus, there is an opportunity to generate other useful quantum states, different from ECSs, by tuning $r$ and $\alpha$ which could give fourther improvements. Similarly, one may consider using approximate even CSS generated in Ref.~\cite{photon_sub02}, instead of using odd CSS obtained by subtracting a single photon \cite{photon_sub01}. In addition, even though it is known that the self Kerr-type non-linear interaction can be performed with small losses in tunable three- or four-level systems \cite{smallLOSS}, it is still an open question as to whether useful higher order non-linear phase operations can also be performed minimizing loss mechanisms.

\vskip 2truecm

\begin{acknowledgements}
We acknowledge J. Dunningham for useful discussions and financial support from the European Commission of the European Union under the FP7 Integrated Project Q-ESSENCE, the FIRST program in Japan, the NRF grant funded by the Korea government (MEST) (No. 3348- 20100018) and the World Class University program.
\end{acknowledgements}

\end{document}